\documentclass[aps,prl,twocolumn,superscriptaddress,floatfix,showpacs,10pt]{revtex4-2}
\usepackage[absolute,overlay]{textpos}
\usepackage{autobreak}
\usepackage{amssymb,amsmath,color,mciteplus,graphicx,subfigure}
\usepackage{calc,epsfig,epstopdf,color,mciteplus,bm,mathrsfs}
\usepackage{times}
\usepackage{float}
\usepackage{lipsum}
\usepackage{amsmath}
\usepackage{soul}
\usepackage[normalem]{ulem}
\usepackage{xcolor}

\usepackage{tabularx}
\usepackage{booktabs}
\usepackage{bbm}
\usepackage{array}
\usepackage{multirow}
\usepackage{amsfonts}
\usepackage{dcolumn}
\usepackage{makecell}
\usepackage{diagbox}
\usepackage{cancel}
\usepackage{placeins}

\usepackage{newtxtext,newtxmath}
\allowdisplaybreaks[4]
\usepackage[unicode=true,
bookmarks=true,bookmarksnumbered=false,bookmarksopen=false,
breaklinks=false,pdfborder={0 0 1},backref=false,colorlinks=true]
{hyperref}
\hypersetup{linkcolor=magenta, urlcolor=blue, citecolor=blue,pdfstartview={FitH}, hyperfootnotes=false, unicode=true}

\newcolumntype{C}[1]{>{\centering\arraybackslash$}p{#1}<{$}}

\begin{document}

\title{Universal Robust Quantum Gates via Doubly Geometric  Control  
}


\author{Hai Xu
}
\affiliation{School of Physical Science and Technology, Guangxi University, Nanning 530004, China}
\affiliation{International Quantum Academy, Shenzhen, 518048, China}

\author{Tao Chen
}
\email{chentamail@163.com}
\affiliation{Key Laboratory of Atomic and Subatomic Structure and Quantum Control (Ministry of Education),\\ Guangdong Basic Research Center of Excellence for Structure and Fundamental Interactions of Matter,\\ and School of Physics, South China Normal University, Guangzhou 510006, China}
\affiliation{Guangdong Provincial Key Laboratory of Quantum Engineering and Quantum Materials, Guangdong-Hong Kong Joint Laboratory of Quantum Matter, and Frontier Research Institute for Physics,\\ South China Normal University, Guangzhou 510006, China}

\author{Junkai Zeng}
\affiliation{International Quantum Academy, Shenzhen, 518048, China}

\author{Xiu-Hao Deng}
\affiliation{International Quantum Academy, Shenzhen, 518048, China}

\author{Fang Gao}
\affiliation{School of Electrical Engineering, Guangxi University, Nanning 530004, China}

\author{\\Xin Wang}
\email{x.wang@cityu.edu.hk}
\affiliation{Department of Physics, City University of Hong Kong, Tat Chee Avenue, Kowloon, Hong Kong SAR, China}

\author{Zheng-Yuan Xue}
\email{zyxue83@163.com}
\affiliation{Key Laboratory of Atomic and Subatomic Structure and Quantum Control (Ministry of Education),\\ Guangdong Basic Research Center of Excellence for Structure and Fundamental Interactions of Matter,\\ and School of Physics, South China Normal University, Guangzhou 510006, China}
\affiliation{Guangdong Provincial Key Laboratory of Quantum Engineering and Quantum Materials, Guangdong-Hong Kong Joint Laboratory of Quantum Matter, and Frontier Research Institute for Physics,\\ South China Normal University, Guangzhou 510006, China}

\author{Chengxian Zhang
}
\email{cxzhang@gxu.edu.cn}
\affiliation{School of Physical Science and Technology, Guangxi University, Nanning 530004, China}
\affiliation{International Quantum Academy, Shenzhen, 518048, China}

\date{\today}

\begin{abstract}

Geometric quantum computation offers a potential route to fault-tolerant quantum information processing by exploiting the global nature of geometric phases. However, achieving controlled high-order suppression of multiple error sources remains a long-standing limitation, particularly in realistic large-scale circuits with complex noise environments. This limitation is largely due to the absence of a general framework that directly characterizes error accumulation and enables systematic improvement. Here we establish such a framework for universal doubly geometric gates by embedding target operations into a hierarchy of “level-$n$” identity constructions. This approach enables direct quantification of error accumulation while removing structural constraints inherent in previous schemes. We analytically show that the defining conditions lead to simultaneous fourth-order suppression of control errors, with a systematic extension to sixth-order suppression via higher-level constructions. Our results establish doubly geometric control as a general and scalable route toward high-order robust quantum gates, with potential implications for fault-tolerant quantum information processing.

\end{abstract}
\maketitle

\textit{Introduction.---} Precise control is essential for scalable quantum information processing \cite{Preskill2018}. Beyond the conventional dynamical phase, geometric phases can arise when a quantum system undergoes cyclic and non-adiabatic evolution in parameter space \cite{Berry_Proc_1984, Wilczek_PRL_1984, Aharonov_1987}. These phases are determined solely by the global geometry of the evolution path and are independent of dynamical details. This property makes geometric phases attractive for quantum computation \cite{Zanardi1999,Pachos1999, Erik_Int_2015, ZhangJiang_2023,liangyan2023,xue2026}, particularly in non-adiabatic geometric quantum computation (NGQC) \cite{WangxbPRL2001, ZhuslPRL2002, ZhuslPRL2003}, where quantum gates are implemented through fast and non-adiabatic evolutions.

Despite these advantages, it remains unclear whether NGQC can provide a practical improvement over dynamical gates  in large-scale quantum circuits \cite{Colmenar_PRXQuantum_2022},
where multiple types of errors are inevitably present. Although geometric phases protect ideal cyclic evolutions, they generally do not constrain how control imperfections accumulate during the evolution. As a result, existing schemes typically suppress only restricted classes of errors or achieve robustness only to low order. Thus, a long-standing challenge in NGQC and other quantum gate strategies is the absence of a general approach for achieving controlled high-order suppression of multiple error sources. Over the past decades, numerous theoretical and experimental studies have shown that NGQC can outperform dynamical gates in suppressing Rabi-type errors \cite{ZhuShiLiang_2003,ZhaoPengZhi_2017,ChenTao_2018,LiuBaoJie_2019,LiKZhao_2020,ZhangChengxian_2020,XuYun_PRL_2020,LianMingJie_PRA_2022,YangXinXin_PRAppl_2023}. However, their performance is degraded in the presence of dephasing noise \cite{DongWenzheng_PRXQ_2021,GuoandXuHai2023,FangZiYu_PRA_2024,ChenTao_PRApplied_2024}. Moreover, their robustness is often assessed using model-dependent fidelity measures \cite{ZhengShiBiao_PRA_2016}, without a unified geometry-compatible metric that directly quantifies error accumulation and enables systematic comparison and optimization. Developing such a framework remains a central open challenge in NGQC.

A key conceptual advance was the recent introduction of the doubly geometric quantum  control (DOG)  protocol \cite{DongWenzheng_PRXQ_2021}.  By combining geometric phases with geometric errors curves \cite{ZengJunkai_2018,JunkaiZeng_2019,Edmunds_2020,Buterakos_2021,EdwinBarnes_2022,Nelson_2023,Deng.24,HaiYongJu2025}, this framework provides a natural route to suppress dephasing errors that limit conventional NGQC. However, despite its appealing geometric formulation, the DOG framework does not readily lend itself to the systematic construction of robust control protocols. In particular, enforcing geometric cyclicity simultaneously in both the geometric phase space and the geometric error-curve space imposes strong constraints on the control landscape, making it difficult to retain universality. As a result, it remains unclear how to systematically design gates that achieve controlled, high-order suppression of multiple noise sources within this framework.

\begin{figure}
	\centering
\includegraphics[width=0.95\columnwidth]{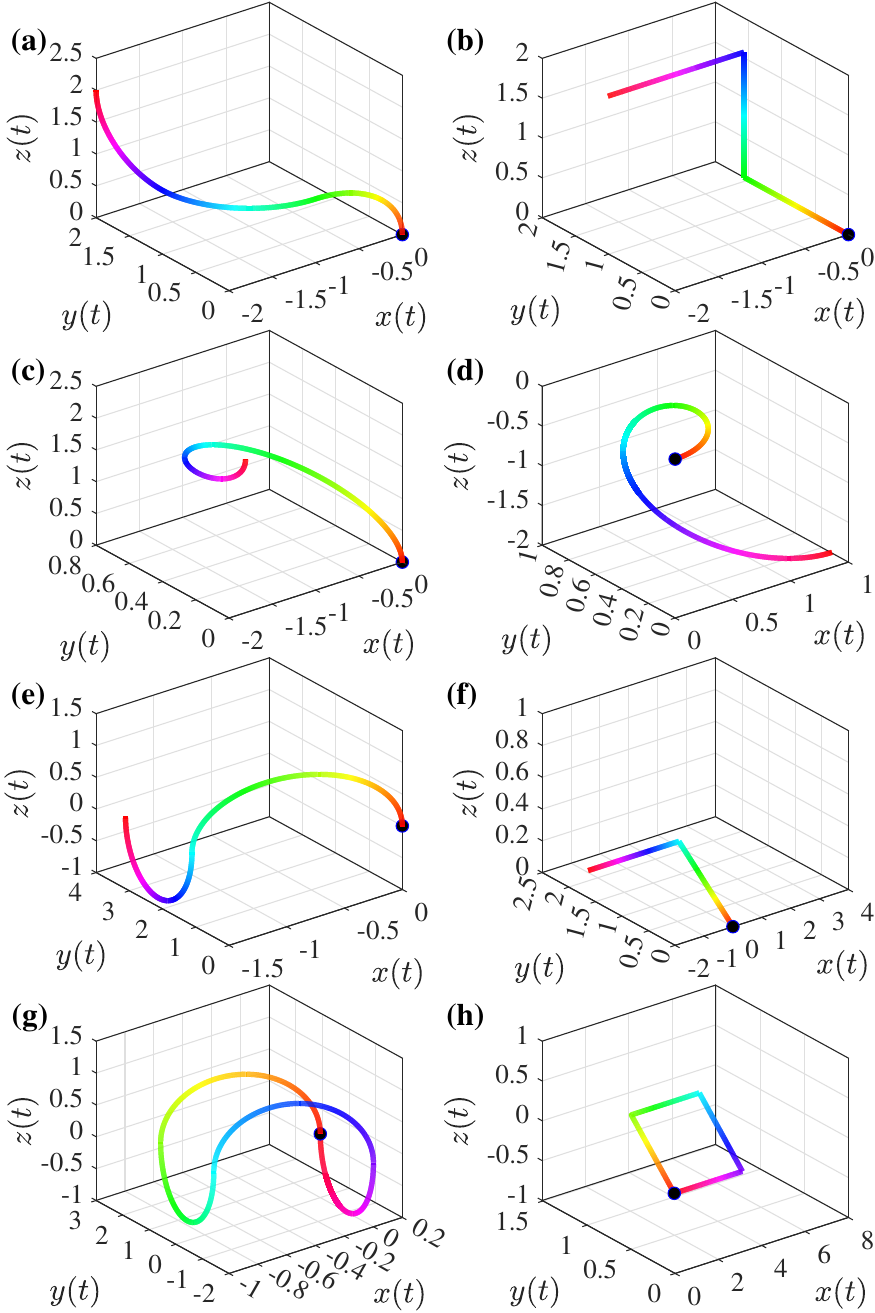}
\caption{Comparison of the $S$ gate error curves for (a) and (b) dynamical scheme \cite{ZhengShiBiao_PRA_2016,SM_UDOG}, (c) and (d) non-cyclic NGQC scheme \cite{ChenTao_PRApplied_2024},
(e) and (f) traditional NGQC scheme \cite{ZhaoPengZhi_2017,XuYun_PRL_2020}, and (g) and (h) level-3 UDOG scheme. The left and right columns correspond to the detuning ($\delta$) and Rabi  ($\epsilon$) errors, respectively. The optimal parameters used here are $\left(\xi_{1}=1.5, \xi_{2}=1\right)$, and the pulse shapes used here are square. Error distances $d$ for panels (a)–(h) are: (a), (b) $\{3.46, 2.72\}$;(c), (d) $\{3.02, 2.26\}$;  (e), (f) $\{3.70, 2.40\}$; and (g), (h) $\{0, 0\}$. }
	\label{Fig:SGate_Errorcurve}
\end{figure}

In this Letter, we present a general analytic framework for constructing universal doubly geometric quantum gates (UDOG) based on a family of parameterized ``level-$n$'' identity operations. For control errors in arbitrary directions, this framework enables direct quantification of error accumulation (Fig.~\ref{Fig:SGate_Errorcurve}) and removes structural constraints inherent in previous DOG schemes, thereby restoring sufficient control flexibility to enforce cyclic evolution simultaneously in both geometric spaces. We show that the defining conditions of UDOG lead to simultaneous fourth-order suppression of two common classes of control errors, while higher-level identity constructions allow a systematic extension to sixth-order suppression (Table.~\ref{table1}). We further verify the performance of UDOG in superconducting transmon qubits and note its broad applicability across physical platforms, as it requires neither external detuning control nor specific constraints on the temporal profile of the driving field. Beyond improving gate robustness, the resulting high-order error suppression provides a potential route to reducing effective physical error rates, which is directly relevant for enhancing the performance of fault-tolerant architectures such as surface codes.

\textit{NGQC in geometric phase space.---}Consider a general two-level control Hamiltonian without introducing an external detuning field ($\hbar=1$)
\begin{equation}
	\mathcal{H}_{c}(t)=\frac{\Omega(t) }{2}[\cos \varphi(t)\sigma_{x}+\sin\varphi(t)\sigma_{y}],
	\label{Eq:H}
\end{equation} 
where $\vec{\sigma}=\left(\sigma_x, \sigma_y, \sigma_z\right)$ is the Pauli vector, $\Omega(t)$ and $\varphi(t)$ are the amplitude and phase dependent on time, respectively. According to the Lewis-Riesenfeld invariant theory \cite{Lewis_Classical_1967,Lewis_exact_1969}, the dynamical invariant $I(t)$ satisfying the von Neumann equation $i\partial{\Pi(t)}/\partial{t}=[\mathcal{H}_{c}(t), \Pi(t)]$ has two eigenstates, i.e., the dressed states $\left|\mu_1(t)\right\rangle=\cos \frac{\theta(t)}{2}|0\rangle+\sin \frac{\theta(t)}{2} e^{i \phi(t)}|1\rangle$ and $ \left|\mu_2(t)\right\rangle=-\sin \frac{\theta(t)}{2} e^{-i \phi(t)}|0\rangle+\cos \frac{\theta(t)}{2}|1\rangle$ \cite{LiuBaoJie_2019,ZhouJian_2021}, where $\theta(t)$ and $\phi(t)$ denote the polar and azimuthal angles on the Bloch sphere, respectively. The state satisfying the Schrödinger equation acquires a global phase factor $|\psi_i(t)\rangle=e^{i f _{i}(t)}|\mu_i(t)\rangle$ ($i=1, 2$), with $f_{i}(t)=\gamma_{d,i}(t)+\gamma_{g,i}(t)$. Here, the dynamical phase is defined as $\gamma_{d,i}(t)= -\int_0^{t}\left\langle\psi_i(t')|\mathcal{H}_{c}(t')| \psi_i(t')\right\rangle d t'$, and $\gamma_{g,i}(t)=f_{i}(t)-\gamma_{d,i}(t)$ is the non-adiabatic geometric phase \cite{Aharonov_1987,SM_UDOG}.  
The geometric gate in the geometric phase space requires two conditions: (i) parallel transport, i.e., the vanishing dynamical phase $\gamma_{d}(t)=0$ or equivalently $\left\langle\psi_i(t)|\mathcal{H}(t)| \psi_i(t)\right\rangle =0$, (ii)  cyclic evolution at the final time $T$, such that $\left|\mu_i(0)\right\rangle=\left|\mu_i(T)\right\rangle$ in the parameter space  \cite{Erik_Int_2015}. Under these conditions, the resulting evolution operator reads $U(T)=\sum_{i}e^{i f_{i}(T)} |\mu_i(0)\rangle \langle \mu_i(0)|$ with $f_{1}(T)=-f_{2}(T)=f(T)$. 

\textit{UDOG in geometric error-curve space.---} 
A quantum system can be subject to various noises. We consider a generic Hamiltonian that models control errors as $\mathcal{H_{\delta}^{\epsilon}}(t)=\epsilon\mathcal{H}_{c}(t)+\delta\sigma_{z}/2$, where $\epsilon$ and $\delta$ represent Rabi and detuning errors, respectively. 
We assume that $\epsilon$ and $\delta$ are unknown stochastic noise terms of small magnitude, so that they can be treated perturbatively and fluctuate slowly compared to the gate duration.
\cite{WangXin_PRA_2014, Nelson_2023,Deng.24}. It is convenient to use the interaction picture, where the perturbed Hamiltonian reads $\mathcal{H}_{I}(t)=U_{c}^{\dagger}(t)\mathcal{H^{\epsilon}_{\delta}}(t)U_{c}(t)$ with $U_{c}(t)=\mathcal{T}e^{-i \int_{0}^{t}\mathcal{H}_{c}(t^{\prime}) d t^{\prime} }$. Direct evaluation of the evolution operator $U_I(t)$ is generally difficult. Instead, we employ the Magnus expansion to the first order \cite{Blanes_PhysRep_2009}
\begin{equation}
	\begin{aligned}
		U_I(t)\simeq\exp \left[-i  A_1(t)\right]=\exp \left[-i  (\epsilon A_{1}^{\epsilon}(t)+\delta A_{1}^{\delta}(t))\right],
	\end{aligned}
	\label{eq:error}
\end{equation}
where
\begin{eqnarray}
A_{1}^{\epsilon}(t)&=& \int_{0}^{t}  U_{c}^{\dagger}(t') \mathcal{H}_{c}(t') U_{c}(t)d t^{\prime} ,\notag\\
A_{1}^{\delta}(t)&=& \frac{1}{2}  \int_{0}^{t} U_{c}^{\dagger}(t') \sigma_z U_{c}(t')  d t^{\prime}.
	\label{eq:At}
\end{eqnarray}
These operators can be expanded in the Pauli basis, i.e.,
\begin{equation}
	\begin{aligned}
		A_{1}^{\epsilon}(t)=\vec{r}^{\epsilon}(t) \cdot \vec{\sigma}=x^{\epsilon}(t) \sigma_x+y^{\epsilon}(t) \sigma_y+z^{\epsilon}(t) \sigma_z,\\
		A_{1}^{\delta}(t)=\vec{r}^{\delta}(t) \cdot \vec{\sigma}=x^{\delta}(t) \sigma_x+y^{\delta}(t) \sigma_y+z^{\delta}(t) \sigma_z.
		\label{eq:rt}
	\end{aligned}
\end{equation} In this way, $\vec{r}^{i}(t)$  ($i=\epsilon,\delta$) can be interpreted as the ``error curve"  in a geometric three-dimensional Euclidean space.  Imposing that the error curves are closed at the final time $T$
\begin{equation}
\begin{aligned}
	\vec{r}^{i}(0)=\vec{r}^{i}(T)=0 \ (i=\delta, \epsilon),
	\label{eq:rt0}
\end{aligned}
\end{equation}
gives $A_{1}^{\epsilon}(T)=A_{1}^{\delta}(T)=0$, and hence $U_{I}(T)=I+\mathcal{O}(\epsilon^{2})+\mathcal{O}(\delta^{2})$.  Consequently, 
the first-order error in the total evolution $U_{\delta}^{\epsilon}=U_{c}(t).U_{I}(t)$ is therefore completely canceled. When the curves are not closed, we can further define the \textit{error distance} $d^{i}=||\vec{r}^{i}(T)-\vec{r}^{i}(0)||$ to quantify the gate error in the $i$-th channel.
As an illustration, Figs.~\ref{Fig:SGate_Errorcurve}(a)–~\ref{Fig:SGate_Errorcurve}(f) display the 3D error curves for the dynamical \cite{ZhengShiBiao_PRA_2016,SM_UDOG}, non-cyclic \cite{ChenTao_PRApplied_2024} and traditional \cite{ZhaoPengZhi_2017,XuYun_PRL_2020} NGQC implementations of the $S$ gate. The nonzero error distances in these schemes indicate that the leading-order errors are not completely canceled.

\textit{Robust UDOG.---}Our general strategy is to enforce cyclic evolution simultaneously in the geometric phase space and the error-curve space, while eliminating dynamical phase accumulation. We first construct a purely geometric operation in the geometric phase space using a pulse sequence in which a $\pi$ rotation is sandwiched between two rotations,

\begin{eqnarray}	\label{eq:U_gs}
U_{c}(\theta_{0},\phi_{0},\gamma_{g}) &=&R_{3}\left(\vartheta_3, \varphi_3 \right) R_{2}\left(\pi, \varphi_2 \right) R_{1}\left(\vartheta_1, \varphi_1 \right)\notag\\
&=&e^{i\gamma_{g} \vec{\sigma} \cdot\vec{n}}.
\end{eqnarray}
Here, $\{ \vartheta_1,  \vartheta_3   \} = \{ \theta_{0}, \pi- \theta_{0} \}$ with $\vartheta_{i}=\int \Omega\left(t\right) d t$, $ \{ \varphi_1,  \varphi_2 ,  \varphi_3   \} = \{ \phi_{0}-\frac{\pi}{2},   \phi_{0}+\gamma_{g}+\frac{\pi}{2},  \phi_{0}-\frac{\pi}{2}\} $ are the piecewise-constant phases in the control Hamiltonian. $\vec{n}=\left(\sin\theta_{0}\cos\phi_{0},\sin\theta_{0}\sin\phi_{0},\cos\theta_{0}\right)$ is the unit vector and $R\left(\vartheta_{i}, \varphi_{i} \right) =\exp \left[-i \vartheta_{i} \left(\cos \varphi_{i} \sigma_{x}+\sin \varphi_{i}\sigma_{y}\right) / 2\right]$  represents an elementary rotation. The $\pi$ rotation is implemented as a parameterized “level-$n$” identity built from an $n$-pulse $\pi$ sequence,

\begin{equation}
	\resizebox{1.04\hsize}{!}{$\left.
	\begin{aligned}
	R_{2}\left(\pi, \varphi_2 \right)&=R_{2}^{n}\left(\pi,\varphi_{2}^{n}\right) R_{2}^{n-1}\left(\pi,\varphi_{2}^{n-1}\right)...
	R_{2}^{2}\left(\pi,\varphi_{2}^{2}\right) R_{2}^{1}\left(\pi,\varphi_{2}^{1}\right).
	\end{aligned} \right.$}
	\label{eq:U_gs2}
\end{equation} For arbitrary tunable phases $\varphi_{2}^{n}$ in the ``level-$n$''  identity, Eq.~\eqref{eq:U_gs} remains a purely geometric operation in the geometric phase space \cite{SM_UDOG}. By choosing these phases appropriately, the corresponding error curves are expected to be closed in the error-curve space, leading to a doubly geometric gate. Then, the next question is how to determine them.

\textit{Level-3 identity UDOG.---} We find that a level-3 identity already provides sufficient flexibility to construct UDOGs and to strongly suppress multiple error channels simultaneously. This is one of the key results of this work \cite{SM_UDOG}. The level-3 identity has the form as
\begin{equation}
R_{2}=R_{2}^{3}\left(\pi,\varphi_{2}^{3} \right)R_{2}^{2}\left(\pi,\varphi_{2}^{2} \right)R_{2}^{1}\left(\pi,\varphi_{2}^{1}\right),
	\label{identity}
\end{equation} where $\{ \varphi_{2}^{1},  \varphi_{2}^{2}, \varphi_{2}^{3}\}=\{ \phi_{0}+\xi_{1}\gamma_{g}+\frac{\pi}{2},  \phi_{0}+\xi_{2}\gamma_{g}+\frac{\pi}{2}, \phi_{0}+\left(1+\xi_{2}-\xi_{1}\right)\gamma_{g}+\frac{3 \pi}{2}\}$.  For a target rotation $U_{c}(\theta_{0},\phi_{0},\gamma_{g})$ in Eq.~\eqref{eq:U_gs}, the parameters $\{\xi_{1},\xi_{2}\}$ remain free and will be fixed by imposing closure of both error curves. To this end, we parameterize the evolution operator as
\begin{equation}
	\begin{aligned}
U_c(t)&=\left(\begin{array}{cc}
	e^{i f(t)} \cos \frac{\theta(t)}{2} & -e^{-i[f(t)+\phi(t)]} \sin \frac{\theta(t)}{2}\\
	e^{i[f(t)+\phi(t)]} \sin \frac{\theta(t)}{2}& e^{-i f(t)} \cos \frac{\theta(t)}{2}
\end{array}\right),
	\label{eq:Ucp}
		\end{aligned} 
\end{equation} and substitute it into Eq.~\eqref{eq:At} to obtain error curves $\vec{r}^{\epsilon}(t)=x^{\epsilon}(t)\hat{x}+y^{\epsilon}(t)\hat{y}+z^{\epsilon}(t)\hat{z}$ and $ \vec{r}^{\delta}(t)= x^{\delta}(t)\hat{x}+y^{\delta}(t)\hat{y}+z^{\delta}(t)\hat{z}$ with
\begin{equation}
		\resizebox{0.9\hsize}{!}{$\left.
			\begin{aligned}
x^{\epsilon}(t) =& -\int_0^t \frac{\Omega(t')}{2} \left(\sin [\varphi(t')-\phi(t')] \sin \left[2 f(t')+\phi(t')\right]\right. \\
&\left.       -\cos\theta(t')\cos\left[\varphi(t')-\phi(t')\right]\cos\left[2 f(t')+\phi(t')\right]\right) d t', \\
y^{\epsilon}(t) = &\int_0^t \frac{\Omega(t')}{2} \left(\sin [\varphi(t')-\phi(t')]  \cos\left[2 f(t')+\phi(t')\right]\right. \\&
\left.+  \cos\theta(t')\cos\left[\varphi(t')-\phi(t')\right]\sin\left[2 f(t')+\phi(t')\right]\right) d t', \\
z^{\epsilon}(t) = &\int_0^t \frac{\Omega(t')}{2} \sin\theta(t')\cos\left[\varphi(t')-\phi(t')\right] d t',
			\end{aligned} \right.$}
		\label{eq:errorp}
	\end{equation}
and
\begin{equation}
		\resizebox{0.8\hsize}{!}{$\left.
			\begin{aligned}
x^{\delta}(t)= & - \frac{1}{2}  \int_0^t  \sin \theta\left(t^{\prime}\right) \cos \left[2 f\left(t^{\prime}\right)+\phi\left(t^{\prime}\right)\right] d t^{\prime}, \\
y^{\delta}(t) = & -\frac{1}{2}  \int_0^t  \sin\theta\left(t^{\prime}\right) \sin \left[2 f\left(t^{\prime}\right)+\phi\left(t^{\prime}\right)\right] d t^{\prime}, \\
z^{\delta}(t)=& \frac{1}{2} \int_0^t \cos \theta\left(t^{\prime}\right) d t^{\prime}.
			\end{aligned} \right.$}
		\label{eq:errorp2}
\end{equation}  The cyclicity condition in the error-curve space requires $x^{i}(T)=y^{i}(T)=z^{i}(T)=0$. Since the parallel-transport condition is enforced, the global phase equals the geometric phase, $f(t)=\gamma_{g}(t)=-\frac{1}{2}\int_{0}^{t}(1-\cos\theta)d\phi$. To eliminate the dynamical phase, we set $\varphi(t)-\phi(t)=\pm\pi/2$ \cite{SM_UDOG}. Then, imposing $\vec{r}^{\epsilon}(T)=0$ yields the following conditions,
\begin{equation}
	\resizebox{1.02\hsize}{!}{$
		\left\{\begin{array}{ll}
			&\left(-(\pi-\theta_0)\sin\left[2 \gamma_g+\phi_0\right]+\theta_0\sin\phi_0+\pi\sin\left[\xi_1 \gamma_g+\phi_0\right] \right.\\
			& \left.-\pi\sin \left[\left(1+\xi_1-\xi_2\right) \gamma_g+\phi_0\right]+\pi\sin \left[\left(2 \xi_1-\xi_2\right) \gamma_g+\phi_0\right]\right)/2 =0;\\ \\
			&\left(-(\pi-\theta_0)\cos\left[2 \gamma_g+\phi_0\right]-\theta_0 \cos\phi_0+\pi\cos\left[\xi_1 \gamma_g+\phi_0\right] \right.\\
			& \left. -\pi\cos \left[\left(1+\xi_1-\xi_2\right) \gamma_g+\phi_0\right]+\pi\cos \left[\left(2 \xi_1-\xi_2\right) \gamma_g+\phi_0\right]\right)/2=0.
		\end{array}\right.$}
	\label{eq:3_conditon1}
\end{equation} 
\begin{table}[tbp]
	\centering
	\tabcolsep=0.14cm
	\renewcommand\arraystretch{1.4}
	\begin{tabular}{cccc}
		\hline\hline
		Types & Detuning error & Rabi error  & References \\
		\hline
		Dynamical  & $1-\frac{3}{2}\delta^{2}$ & $1-\frac{3}{32}\pi^2\epsilon^2$ &\cite{SM_UDOG,ZhengShiBiao_PRA_2016} \\
		Non-cyclic NGQC  &$1-\frac{83}{73}\delta^2$  &$1-\frac{7}{108}\pi ^2\epsilon^2$ & \cite{ChenTao_PRApplied_2024} \\
		Traditional NGQC  &$1-(1+\frac{1}{\sqrt{2}})\delta^2$  &$1-\frac{(2-\sqrt{2})}{8}\pi ^2\epsilon^2$ & \cite{ZhaoPengZhi_2017,XuYun_PRL_2020} \\
		Level-3 UDOG & $1-(1-\frac{1}{\sqrt{2}}) \delta^4$ & $1-\frac{(2-\sqrt{2})}{32}\pi^{4} \epsilon^{4}$ & This work \\
		Level-5 UDOG & $1-(1+\frac{1}{\sqrt{2}})\delta^6$ & $1-\frac{(2-\sqrt{2})}{128}\pi ^6\epsilon^6$ & This work \\
		\hline\hline
	\end{tabular}
	\caption{The robustness comparison with the $S$ gate fidelity of dynamical scheme, traditional NGQC scheme, and our level-3 and 5 identity UDOG.}
	\label{table1}
\end{table} 
While $\vec{r}^{\delta}(T)=0$ gives 
\begin{equation}
	\resizebox{0.96\hsize}{!}{$
		\left\{\begin{array}{ll}
			&-\cos^{2}\frac{\theta_{0}}{2}\cos \left[2 \gamma_g+\phi_0\right] - \cos \left[\xi_1 \gamma_g+\phi _0\right]-\sin^{2}\frac{\theta_{0}}{2} \cos\phi_0 \\
			&+\cos \left[\left(1+\xi_1-\xi_2\right) \gamma_g+\phi_0\right]+\cos \left[\left(2 \xi_1-\xi_2\right) \gamma_g+\phi_0\right]=0;\\ \\
			&\cos^{2}\frac{\theta_{0}}{2} \sin\left[2 \gamma_g+\phi_0\right]+\sin \left[\xi_1 \gamma_g+\phi_0\right]-\sin^{2}\frac{\theta_{0}}{2} \sin\phi_0\\
			& -\sin \left[\left(1+\xi_1-\xi_2\right) \gamma_g+\phi_0\right]-\sin \left[\left(2 \xi_1-\xi_2\right) \gamma_g+\phi_0\right]=0.
		\end{array} \right.$}
	\label{eq:3_conditon2}
\end{equation} 
For any desired rotation specified by $(\theta_{0},\phi_{0},\gamma_{g})$, solving Eqs.~\eqref{eq:3_conditon1} and \eqref{eq:3_conditon2} determines $\xi_{1}$ and $\xi_{2}$, thus constructing a doubly geometric operation. In particular, the above derivation does not impose any restriction on the pulse shape $\Omega(t)$, which can be either time-independent or time-dependent \cite{SM_UDOG}. As an example, $z$-axis rotation corresponds to $\theta_{0}=\phi_{0}=0$. Solving Eq.~\eqref{eq:3_conditon1} yields $\xi_1=1.5$ and $\xi_2=1$
(which also satisfy Eq.~\eqref{eq:3_conditon2}) for arbitrary $\gamma_{g}$, confirming the UDOG condition for both error channels. Figs.~\ref{Fig:SGate_Errorcurve}(g) and \ref{Fig:SGate_Errorcurve}(h) show the resulting error curves for the UDOG $S$ gate ($\gamma_g=-\pi/4$), where the error distances vanish for both errors. The construction conditions of other gates can be found in Ref. \cite{SM_UDOG}.

\begin{figure}
	\centering
\includegraphics[width=1\columnwidth]{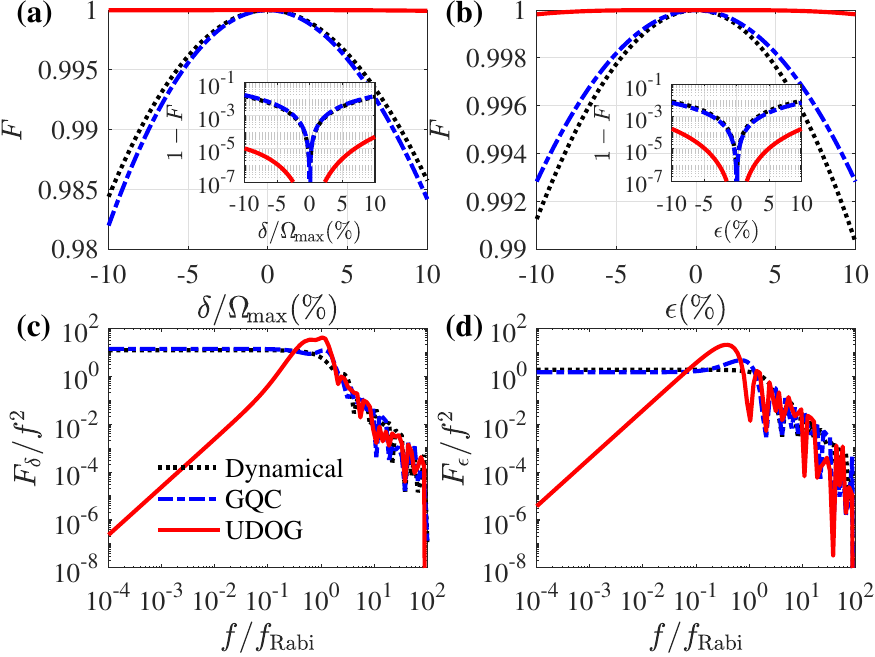}
\caption{Gate fidelity of the level-3 $S$ gate as a function of (a) detuning error $\delta$ and (b) Rabi error $\epsilon$. The optimal parameters are $\left(\xi_{1}=1.5,\xi_{2}=1\right)$, and the pulse shapes used here are square.  (c) and (d) represent the corresponding filter functions \cite{Green.13}.}
	\label{Fig:SGate_Robustness}
\end{figure}

\textit{Robustness verification.---} To evaluate the performance of the level-3 UDOG, we derive the gate fidelity as \cite{SM_UDOG}
\begin{equation}
F	\simeq 1-\frac{\beta^{2}}{4}\sum_{k=1,m=1}^{2} |D_{km}\left(t\right)|^{2}  +O\left(\beta^{4}\right),
		\label{Eq:Fidelity}
\end{equation} where $D_{km}\left(t\right)\equiv\int_{0}^{t}\langle\psi_{k}(t')| V(t')|\psi_{m}(t')\rangle dt'$, and $\beta V(t)$  is a general error Hamiltonian with $\beta$ a small parameter. We first consider the Rabi-error case with $\beta=\epsilon$, for which the error Hamiltonian is $\beta V_{\epsilon}(t)=\epsilon\mathcal{H}_{c}(t)$.
We obtain 
 \begin{equation}
\begin{aligned}
	&D_{11}^{\epsilon}\left(t\right)=\int_{0}^{t} \frac{\Omega(t')}{2}\sin\theta\left(t'\right)\cos\left[\varphi (t')-\phi(t')\right] d t^{\prime},\\
	&D_{21}^{\epsilon}\left(t\right)=\int_{0}^{t} \frac{\Omega(t')}{2}\left(\cos\theta\left(t'\right)\cos\left[\varphi (t')-\phi(t')\right]\right. 
	\\&
	\left.\quad\quad\quad\quad+i\sin\left[\varphi (t')-\phi(t')\right]\right)e^{i\left(2f(t')+\phi (t')\right)} d t'.
\end{aligned}
\label{Eq:f_Rabi}
\end{equation}   Comparing Eq.~\eqref{Eq:f_Rabi} with  Eq.~\eqref{eq:errorp}, we find
\begin{equation}
\begin{aligned}
{\rm{Re}}\left(D_{21}^{\epsilon}\left(t\right)\right)=x^{\epsilon}(t),
{\rm{Im}}\left(D_{21}^{\epsilon}\left(t\right)\right)=y^{\epsilon}(t),
	D_{11}^{\epsilon}\left(T\right)=z^{\epsilon}(t).
\end{aligned}
\label{Eq:Rabicorrespondence}
\end{equation} For the detuning-error case with $\beta=\delta$, the corresponding error Hamiltonian is $\beta V_{\delta}\left(t\right)=\frac{\delta}{2}\sigma_{z}$.
We obtain
\begin{equation}
\begin{aligned}
	&D_{11}^{\delta}\left(t\right)= \int_{0}^{t}\frac{1}{2}\cos\theta\left(t^{\prime}\right) d t^{\prime},\\
	&D_{21}^{\delta}\left(t\right)=-\int_{0}^{t} \frac{1}{2}\sin\theta\left(t^{\prime}\right)e^{i\left(2f\left(t^{\prime}\right)+\phi\left(t^{\prime}\right)\right)} d t^{\prime}.
\end{aligned}
\label{Eq:D21_UDOG_Detuning2}
\end{equation}
Comparing  Eq.~\eqref{Eq:D21_UDOG_Detuning2} with Eq.~\eqref{eq:errorp2}, we find
\begin{equation}
\begin{aligned}
{\rm{Re}}\left(D_{21}^{\delta}\left(t\right)\right)=x^{\delta}(t),
{\rm{Im}}\left(D_{21}^{\delta}\left(t\right)\right)=y^{\delta}(t),
	D_{11}^{\delta}\left(t\right)=z^{\delta}(t).
\end{aligned}
\label{Eq:Detucorrespondence}
\end{equation} 
Eq.~\eqref{Eq:Rabicorrespondence} and ~\eqref{Eq:Detucorrespondence} show that the UDOG condition $x^{i}(T)=y^{i}(T)=z^{i}(T)=0 \ (i=\delta, \epsilon)$ simultaneously suppresses both Rabi and detuning errors to the fourth order.

We further find that fidelity can be substantially enhanced using a level-5 identity \cite{SM_UDOG}. As shown in Table.~\ref{table1}, expanding the fidelity in powers of  $\epsilon$ or $\delta$ reveals that the level-5 UDOG achieves sixth-order error dependence. This is another key result of this work. In Figs.~\ref{Fig:SGate_Robustness}(a) and~\ref{Fig:SGate_Robustness}(b), we compare the robustness of UDOG with dynamical \cite{ZhengShiBiao_PRA_2016} and traditional NGQC \cite{ZhaoPengZhi_2017,XuYun_PRL_2020} schemes. The fidelity is defined as $\operatorname{Tr}[ U_{\rm{ideal}}^{\dagger}\cdot U^{\epsilon/\delta}] / 2$, where $U^{\epsilon/\delta}$ is the error-affected evolution operator.  Across the entire considered regime, the UDOG fidelity remains nearly unchanged. Figs.~\ref{Fig:SGate_Robustness}(c) and~\ref{Fig:SGate_Robustness}(d) show the filter functions \cite{Green.13} for the three schemes, revealing a pronounced reduction of low-frequency components and superior resilience to quasi-static detuning and Rabi noise  in the UDOG case. This advantage is most pronounced when the environmental spectrum is dominated by $1/f$-type or quasi-static noise, a regime relevant to superconducting and spin-qubit platforms. 

\begin{figure}
	\centering
	\includegraphics[width=1\columnwidth]{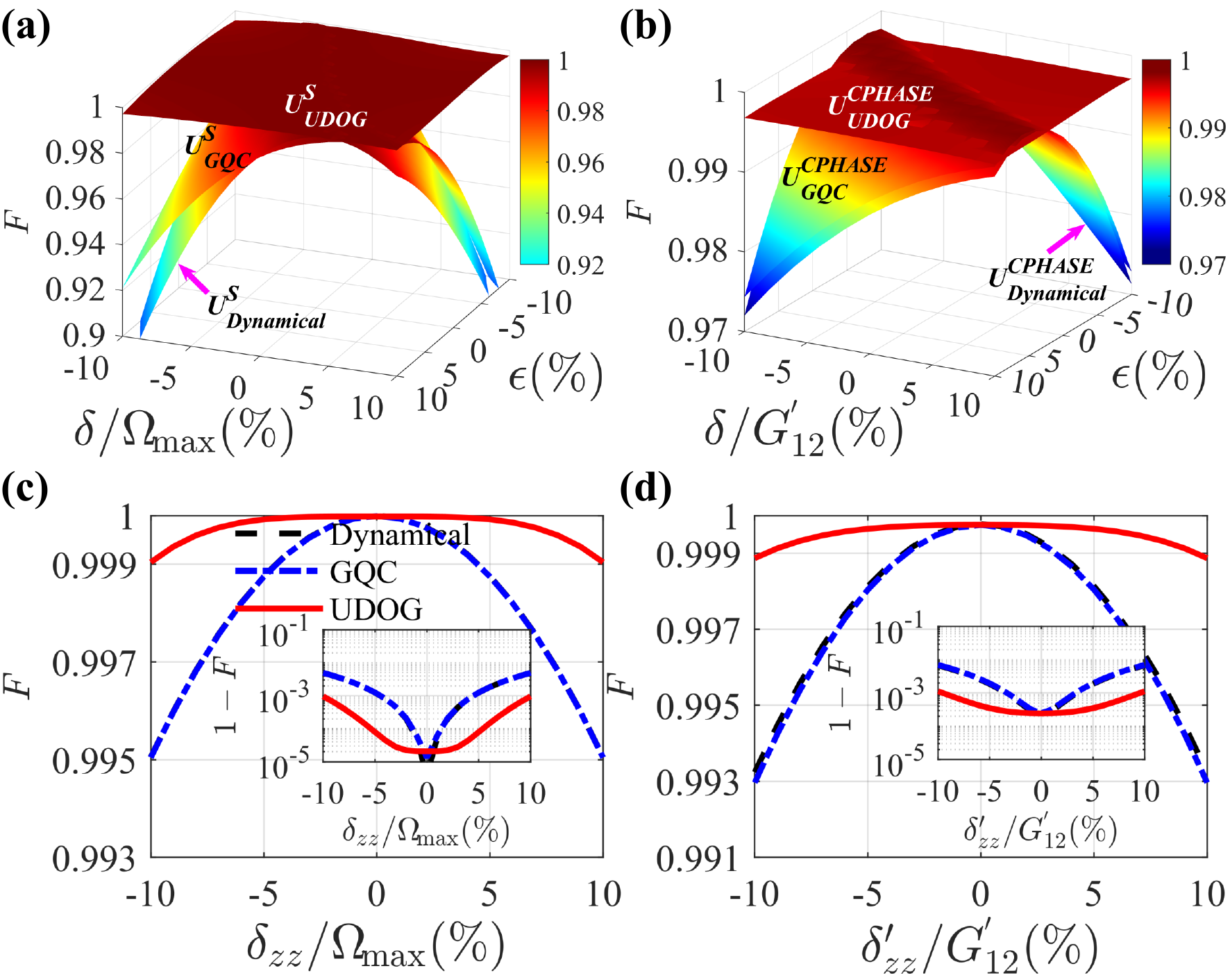}
	\caption{Gate fidelity  in the superconducting transmon qubits. (a) corresponds the level-3 identity $S$ gate, and (b) the CPHASE gate, where the tunable parameters used here are $\left(\xi_{1}=1.5,\xi_{2}=1\right)$.  (c) and (d) correspond to the $X$ and iSWAP gate considering the  $ZZ$ crosstalk error $\delta_{zz}$ ($\delta_{zz}^{\prime}$), where the  tunable parameters used here are $\left(\xi_{1}=-5/3,\xi_{2}=5/3\right)$.}
	\label{Fig_Ssimu_Xcrosstalk}
\end{figure}

\textit{A Physical Implementation.---} Many qubit platforms are promising for verifying the noise-resilient feature of doubly geometric gates, such as superconducting circuits  \cite{DevoretMH_2013,krantzP_2019,KjaergaardMorten_2020} and silicon-based quantum dots \cite{Burkard_RevModPhys_2023}. Here, we take the superconducting transmon qubit as an example and consider the level-3 identity  \cite{SM_UDOG}. 
In transmon qubits, various types of error can occur. 
In particular, the Rabi error is induced by inaccurate control in the driving field, and the detuning error arises from qubit-frequency drift due to flux noise \cite{krantzP_2019}. 
To verify the robustness of doubly geometric operation, we analyze these errors through contour plots using two representative gates: a single-qubit $S$ gate (Fig.~\ref{Fig_Ssimu_Xcrosstalk}(a)) and a controlled-phase (CPHASE) gate (Fig.~\ref{Fig_Ssimu_Xcrosstalk}(b)).  Clearly, the UDOG scheme can significantly enhance the gate performance compared to both dynamical and traditional NGQC schemes. Here, we set a moderate decoherence rate of $\kappa=2\pi\times3 \ \mathrm{kHz}$ \cite{KjaergaardMorten_2020}. Another critical error source is the residual $ZZ$ crosstalk between qubits \cite{KandalaA_PRL_2021,NiZhongchu_PRL_2022,WeiKX_PRL_2022}, with an interaction strength of $\eta_{zz}^{ij}$, 
which is a major challenge for scaling superconducting quantum devices. 
We further investigate $X$ and iSWAP gates as test cases, 
as shown in Figs.~\ref{Fig_Ssimu_Xcrosstalk}(c) and~\ref{Fig_Ssimu_Xcrosstalk}(d), 
our UDOG can effectively suppress infidelity induced by $ZZ$ crosstalk, and substantially outperform other 
schemes. The detailed physical implementation is presented in Ref.~\cite{SM_UDOG}.

\textit{Conclusion.---} We have established a general analytic framework that clarifies the structural origin of robustness in geometric quantum gates. Within this framework, we show that high-order error suppression emerges as a direct consequence of geometric constraints through ``level-$n$'' identity constructions. More broadly, our results highlight the potential of UDOG for robust quantum information processing. In fault-tolerant architectures such as surface codes~\cite{Fowler.12}, 
improving the scaling of physical errors from 
second to higher orders effectively reduces the physical error rate at fixed control precision. This can be interpreted as an enhancement of the effective code distance, leading to a substantial suppression of logical errors without increasing hardware overhead. In addition, the geometry-based protection offered by our scheme is promising for other applications ranging from noise-resilient phase estimation \cite{LouzonDaniel.PRL25} to NISQ algorithms such as variational quantum eigensolver algorithms \cite{JulesTilly.22} and quantum approximate optimization algorithms \cite{KostasBlekos2024}.

\bigskip
We thank Wen-Zheng Dong for fruitful  discussions. This work was supported by the National Natural Science
Foundation of China (Grant No. 11905065, No. 62171144, No. 92576110, No. 12474489,
and No. 12275090), the Shenzhen International Quantum Academy (Grant No. SIQA2025KFKT08), the Shenzhen Fundamental Research Program (Grant No. JCYJ20240813153139050) and the Guangdong Provincial Quantum Science Strategic Initiative (Grant No. GDZX2203001, GDZX2403001).. 


\let\oldaddcontentsline\addcontentsline     
\renewcommand{\addcontentsline}[3]{}


%

\end{document}